\documentclass{aa}
\usepackage{graphics}
\usepackage{natbib}

\begin{document}
\def\ab{$\sim$}
\def\frac{$''$\hspace*{-.1cm}}
\def\deg{$^{\circ}$\hspace*{-.1cm}}
\def\min{$'$\hspace*{-.1cm}}
\def\h2{H\,{\sc ii}}
\def\hi{H\,{\sc i}}
\def\hb{H$\beta$}
\def\ha{H$\alpha$}
\def\hd{H$\delta$}
\def\heii{He\,{\sc ii}}
\def\hg{H$\gamma$}
\def\sii{[S\,{\sc ii}]}
\def\siii{[S\,{\sc iii}]}
\def\oiii{[O\,{\sc iii}]}
\def\oii{[O\,{\sc ii}]}
\def\hei{He\,{\sc i}}
\def\sm{$M_{\odot}$}
\def\slum{$L_{\odot}$}
\def\mdot{$\dot{M}$}
\def\x{$\times$}
\def\sec{s$^{-1}$}
\def\cm2{cm$^{-2}$}
\def\mcube{$^{-3}$}
\def\lam{$\lambda$}

\def\ac{Al~{\sc{iii}}\ }
\def\cb{C\,{\sc{ii}}}
\def\cc{C\,{\sc{iii}}}
\def\cd{C\,{\sc{iv}}}
\def\ca{Ca\,{\sc{ii}}}
\def\crb{Cr\,{\sc{ii}}}
\def\feb{Fe\,{\sc{ii}}}
\def\fec{Fe\,{\sc{iii}}}
\def\hea{He\,{\sc{i}}}
\def\heb{He\,{{\sc ii}}}
\def\nb{N\,{\sc{ii}}}
\def\nc{N\,{\sc{iii}}}
\def\nd{N\,{\sc{iv}}}
\def\ne{N\,{\sc{v}}}
\def\na{Na~{\sc{i}}\ }
\def\nf{Ne~{\sc{i}}\ }
\def\soc{S~{\sc{iii}}\ }
\def\sd{S\,{\sc{iv}}}
\def\sib{Si\,{\sc{ii}}}
\def\sic{Si\,{\sc{iii}}}
\def\sid{Si\,{\sc{iv}}}
\def\tib{Ti\,{\sc{ii}}}

\title{Tight LMC massive star clusters R\,127 and R\,128\thanks
   {Based on observations obtained at the European Southern 
   Observatory, La Silla, Chile}
}

\offprints{Fr\'ed\'eric Meynadier, Frederic.Meynadier@obspm.fr}

\date{Received 12 November 2002 / Accepted 15 January 2003}

\titlerunning{LMC R\,127/R\,128}
\authorrunning{Heydari-Malayeri et al.}

\author{M. Heydari-Malayeri\inst{1} \and F. Meynadier\inst{1}
    \and Nolan R. Walborn\inst{2}
}

\institute{{\sc lerma}, Observatoire de Paris, 61 Avenue de l'Observatoire, 
F-75014 Paris, France \and Space Telescope Science Institute, 3700 San Martin
Drive, Baltimore, Maryland 21218, USA }

\abstract{We study the Large Magellanic Cloud (LMC) star clusters R\,127 
and R\,128 using imaging and spectroscopy obtained at the ESO NTT
telescope. An advanced image restoration technique allows us to
resolve these two clusters into at least 14 and 33 stars
respectively and obtain their photometry. In particular, we show that
the core of R\,127 is composed of at least four stars and identify the Luminous
Blue Variable (LBV) component. The closest neighbor of the LBV (star
\#8) is 1\frac.5 away.  Moreover, from medium dispersion spectroscopy
we determine the spectral types for 19 stars in and near both
clusters, and in particular present the first spatially resolved
observation of the second brightest component of the R\,127 cluster
(star \#3) situated 3\frac.3 from the LBV. By comparing with
evolutionary models we also  look into the stellar ages. The
oldest stars of the cluster are  \ab\,6--8 Myr old, whereas the most
massive star of the region (\#7), formed \ab\,3 Myr ago as an  80\,\sm\,
star, has turned into an LBV, the  ``R\,127'' star. \\
}

\maketitle

\hspace*{-.7cm} 

\section{Introduction}

\object{R\,127} and \object{R\,128} 
\citep{feast} are two Large Magellanic Cloud (LMC)
massive stars situated some 300 pc south of the famous 30 Doradus star
forming factory and 150 pc west of the \h2 region N158
\citep{henize}. They are in fact the brightest members of the two
adjacent tight clusters forming NGC\,2055 towards the center of the OB
association LH\,94 \citep{lh}. Although they lie in the \h2 region
DEM\,248 \citep{Davies1976}, they are not apparently linked to any
specific nebulosity.\\

R\,127, also called \object{Sk--69$^{\circ}$\,220} \citep{Sanduleak1970} or
\object{HDE\,269858}, is recognized as a Luminous Blue Variable (LBV), 
 a very rare class of evolved massive stars.
\citet{parker} listed 8 members in the LMC, 
while \citet{gend01} increased the number of members and candidates to
21.  In the past they were called S Dor variables, a designation
introduced by \citet{kukarkin}, and there is a strong tendency by some
workers to use the original name.
These most luminous stars ({\it log} $L/L_{\odot}$\,\ab\,5.0--6.3) are 
characterized by irregular photometric and spectral variations over 
decades and evolve from a hot (OB-type) visual minimum phase to a cooler 
(A-type), visual maximum
\citep{humph94}. In the Hertzprung-Russell (H-R) diagram they are located
very close to the observed upper luminosity boundary for very massive
stars, the  Humphreys-Davidson limit. LBVs are characterized
by extreme instability, dramatic outbursts and high mass loss. Between
violent eruptions LBVs still lose mass at high rates, 10$^{-5}$ to
10$^{-4}\,$\sm\,yr$^{-1}$ (\citet{lam01} and references therein).  The
material ejected (up to several \sm) forms a nebula around LBVs with
typical sizes of 0.5--2 pc \citep{nota97}.  The properties of LBVs are
reviewed by \citet{humph94} and more recently in the proceedings of a
special workshop edited by \citet{notlam}.  LBVs are believed to be the
precursors of Wolf-Rayet stars; however their precise evolutionary
state is still poorly understood. R\,127 is quite noteworthy even
among this small population; it is the brightest LBV star in the LMC
($M_{bol}=-10.5$ mag) and has the largest magnitude variation
($\Delta$V\,\ab\,2.5 mag) \citep{stahl83,humph94,gend01}.  Due to its
exceptional characteristics, R\,127 has been the subject of much research
in the past.  R\,127 was classified as OIafpe extr or WN\,9-10
by \citep{walborn77, walborn82}. A
brightening of 0.75\,mag later caused its classification 
 as an S Dor
variable \citep{stahl83}. It was then included in the Long-Term
Photometry of Variables  organized by
\citet{sterken83}, and published in several papers \citep{manfroid91, 
manfroid94, sterken93, sterken95, genderen97}.
The monitoring showed that R\,127 had
a very bright phase around the end of 1986
with $y=9.05$ mag, and as such  
was the brightest star in the LMC
\citep{wolf88}, only surpassed in February 1987 by SN1987A.
R\,127 reached its visual
maximum of $V$\,\ab\,8.8 mag around the year 1989 
and then slowly
decreased to $V$\ab\,9.4 mag in \ab\,1995. A new maximum of
$V$\,\ab\,9.3 mag was attained in 1997 coinciding with the time of
present observations. Subsequently it became fainter, with $V$\,\ab
10.4 mag, around the year 2000 (Stahl 2002, private communication). \\

Infra-red observations have shown that R\,127 is surrounded by a dust shell
\citep{stahl84}. As this dust shell is believed to be a
crucial element for understanding the evolution of massive stars,
it has been extensively studied using imagery
\citep{stahl87}, coronography \citep{clampin},
spectroscopy \citep{smith}, and polarimetry
\citep{schulte}.\\

R\,128, otherwise \object{Sk--69$^{\circ}$\,221} or
\object{HDE\,269859}, is a supergiant B2 Ia
\citep{fitz91}. It is variable with a total range of 0.32 
mag between 1983 and 1990, which is very large for  its 
spectral type \citep{gend98}. Interestingly, this star has
formerly been considered as an LBV candidate \citep{gend01}. \\

The studies so far devoted to these two interesting bright stars have
mainly dealt with their individual characteristics. However, recent
findings, from both high-resolution observations and theoretical
works, suggest that massive stars form in groups.  
Therefore, knowing the characteristics of the cluster members is
necessary for better understanding the formation and evolution of
these brightest, probably the most massive, stars of the group.  
\citet{notaIAU143} looked into the multiplicity of R\,127 by
obtaining high-resolution, ground-based images with the STScI
coronograph and reported the presence of 20 stars towards R\,127.
Although these components are mainly field stars detached from
R\,127 proper, they detect a relatively close component lying 3\frac.5
north-west of the bulk of R\,127. \\

 The present work is therefore devoted to the photometry and
 spectroscopy of the cluster members. Using  high resolution
 imaging techniques, we aim at resolving the cores of the clusters R\,127 and
R\,128. Moreover, high-spatial-resolution spectroscopy will allow us 
 to study the physical properties of so far unknown members of
 these two tight clusters.\\

\section{Observations and data reduction}
\subsection{Sub-arcsecond imaging and deconvolution photometry}

The R\,127 and R\,128 clusters were observed on 20 November 1997 using
the ESO New Technology Telescope (NTT) equipped with the active optics
SUperb Seeing Imager (SUSI). The detector was a Tektronix CCD (\#42)
with $1024\times1024$ pixels of 24 $\mu$m (0\frac.13 on the sky), and
the seeing varied between 0\frac.81 and 1\frac.34\, ({\sc fwhm}). \\

The observations were performed in the $uvby$ Str\"omgren photometric
system using the ESO filters \# 715, 716, 713, and 714 respectively.
We were particularly careful to keep most of the brightest stars in
the field under the detector's saturation level to have at our
disposal high quality Point Spread Function (PSF) stars. This led us
to adopt exposure times of 350, 175, 210 and 140 seconds in $u$, $v$,
$b$ and $y$ respectively.  We also used ditherings of
5\frac\,--10\frac\, for bad pixel rejection and in order to be able to
use the full oversampling capabilities of the MCS deconvolution
algorithm \citep{Magain98}.  Indeed when performing simultaneous
deconvolution of several frames, the algorithm uses the different
frame centerings as a constraint while decreasing the pixel size.  We
took a grid of seven dithering positions for each filter.  Luckily,
the targets of interest are close enough to be contained in a single
SUSI field of view. Unfortunately, the $u$ images could not be used
for the photometry due to their insufficient quality.\\

\begin{figure*}
\begin{center}
\resizebox{18cm}{!}{\includegraphics{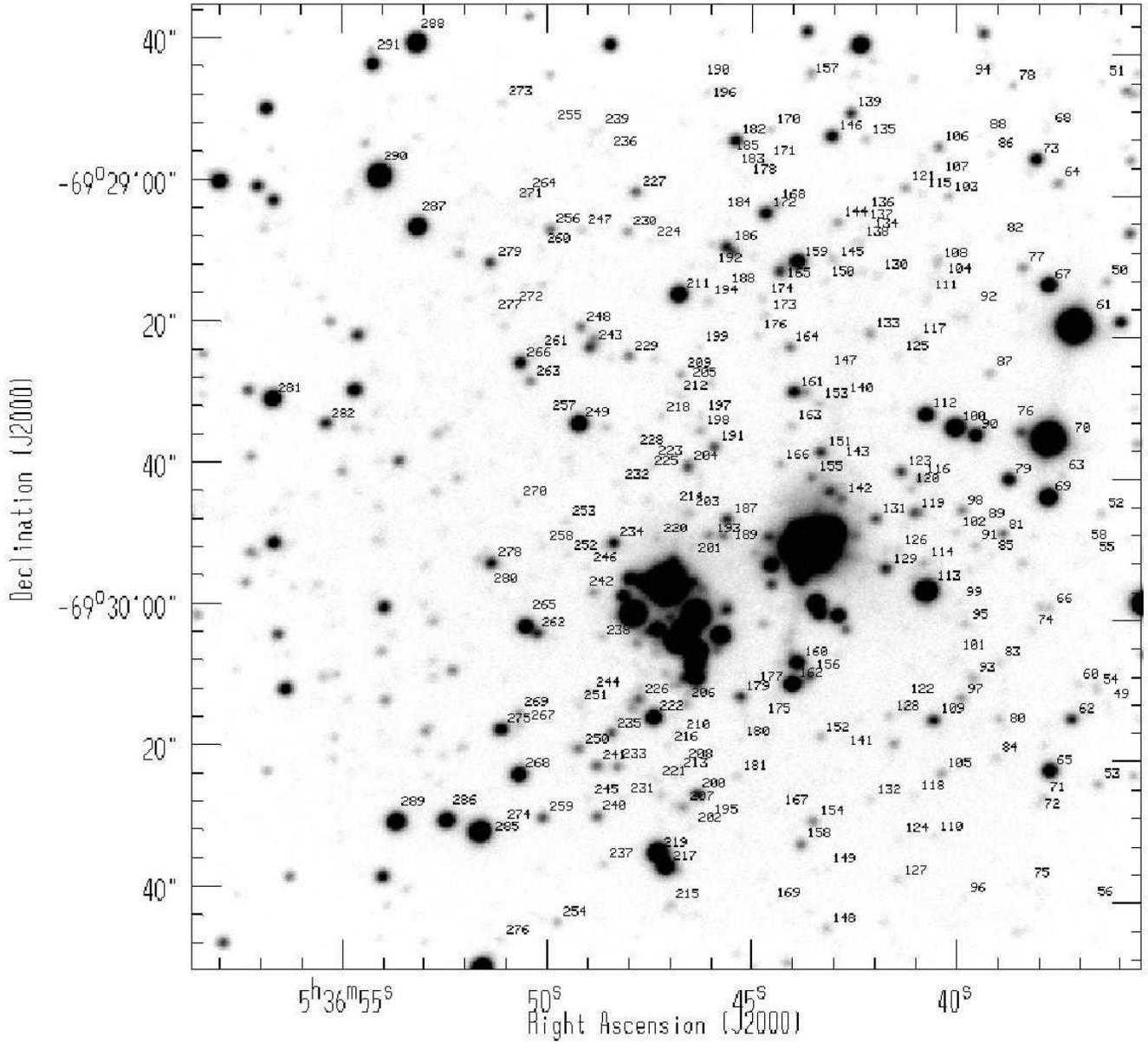}}
\caption{The LMC star clusters R\,127 and R\,128 as seen through the 
Str\"omgren $v$ filter. The image is based on co-adding 7 dithered
individual exposures. R\,127 is the cluster lying to the right of the
image center. The field size is 2\min\,\,\x\,2\min, corresponding to 31
pc \x\, 31 pc. North is up and east to the left.
\label{global}} 
\end{center}
\end{figure*}

Photometry was derived in the Str\"omgren $v$, $b$ and $y$ filters
according to the following procedure: after bias subtraction and
flat-fielding, the seven frames were co-added in each of the filters.
The photometry of the stars lying outside  the compact clusters
was performed on the resulting frames through the DAOPHOT reduction 
package. This yielded the photometry of 233 stars situated outside clusters
R\,127 and R\,128 (Fig.~\ref{global}).\\

The crowded clusters were processed with the MCS restoration
algorithm.  The deconvolution was performed on a 256\,\x\,256 pixel
region containing both R\,127 and R\,128.  The MCS code, proposed and
implemented by \citet{Magain98}, results from a new approach to
deconvolution taking care not to violate the \citet{Shannon1949}
sampling theorem: the images are deliberately not deconvolved with the
observed PSF, but with a narrower function, chosen so that the final
deconvolved image can be properly sampled, whatever sampling step is
adopted to represent the final data.  For this purpose, one chooses
the final, well-sampled PSF of the deconvolved image and computes the
PSF which should be used to perform the deconvolution.  The observed
PSF is constructed from several stars close enough to the clusters in
order to avoid any possible PSF variation across the field. \\

\begin{figure*}
\begin{center}
\resizebox{15cm}{!}{
\includegraphics{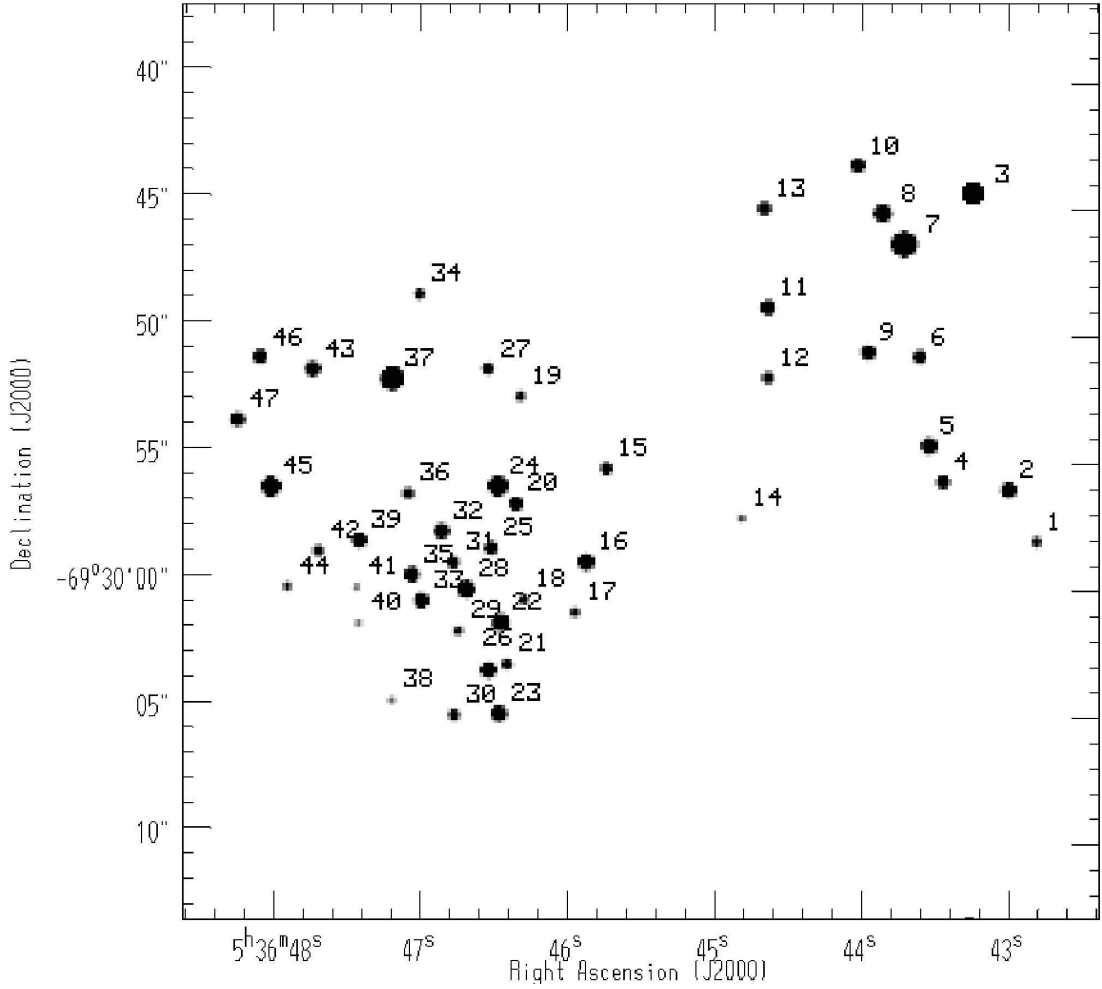}
}
\caption{ Restored Str\"omgren $y$ image of the clusters R\,127 and R\,128 
using the MCS deconvolution method. Star \#7 is the main component of
the R\,127 cluster, resolved into 14 components. R\,128 is shown to be
made of at least 33 components. Stars outside the clusters were masked
during the deconvolution process.  The field size is 33\frac.3 \x\,
33\frac.3 (8.3 pc \x\ 8.3 pc). North is up and east to the left.
\label{dec} 
} 
\end{center}
\end{figure*}

The restoration resolves the  R\,127 and R\,128 clusters
into 14 and 33 components respectively (Fig.~\ref{dec}). The
deconvolution method requires a high S/N ratio for the sources in
order to achieve accurate photometry and astrometry. Therefore, very
faint stars too close to the bright components were excluded from the
deconvolution process. Star \#13 was detected but its magnitude
could not be accurately measured, due to the presence of
relatively strong process residuals. \\

A technical problem during the observing run prevented us from
obtaining adequate standard star observations necessary for
calibrating the photometry.  The available calibration frames 
allowed us however to correct for the atmospheric extinction 
for all filters and derive the zero point for the $y$  filter.  
Results concerning $y$ magnitudes were checked using the photometric 
data available in the literature for three stars in the field, 
 Sk--69$^{\circ}$\,217, Sk-69$^{\circ}$\,218 and R\,128 
\citep{ardeberg, isser75}. Intrinsic $(b-y)_0$ colors were calculated 
for each star of known spectral type (see Sect. 4), in two steps: 
first we established their effective temperatures from the  
calibration by \citet{Vacca96}, then the corresponding theoretical 
$(b-y)_0$ colors were deduced using \citet{Relyea78}. An average  
interstellar reddening of $E(B-V)=0.15$ mag
\citep{st-louis}, or 
$E(b-y)=0.12$ mag \citep{kaltcheva}, was used to obtain 
the observational  $(b-y)$ colors. \\

The final photometric results for the two clusters R\,127, R\,128, and
the brightest field stars are presented in Tables \ref{photor127},
\ref{photor128}, and \ref{photobrightest} which also list the
$\alpha$-$\delta$ (2000.0) positions of the stars. The astrometry was
obtained from the identification of several bright, isolated stars
with GSC 2.2 stars. The positional accuracy is found to be around
2\frac.\\

A word of caution seems appropriate regarding the colors.  The
powerful deconvolution method has allowed us to resolve the compact
clusters R\,127 and R\,128 revealing their so far unknown components.
Furthermore, the code has enabled us to perform the photometry of the
tight components.  However, we should underline that this photometry
is relative for a number of reasons which have nothing to do with the
limitations of the code.  The shortcomings in the observation of
standard photometric stars particularly affect the colors, and we
have therefore taken care not to over-interpret them. \\

\subsection{Spectroscopy with NTT/EMMI}

The EMMI spectrograph attached to the ESO NTT telescope was used on 22
November 1997 (BLMD mode) to obtain several long slit spectra.  The
grating was \#\,12 centered on 4350\,\AA\, and the detector was a
Tektronix CCD (\#\,31) with 1024$^{2}$ pixels of size 24 $\mu$m.  The
range was 3810--4740\,\AA\, and the dispersion
38\,\AA\,mm$^{-1}$, giving {\sc fwhm} resolutions of $2.70\pm0.10$
pixels or $2.48\pm0.13$\,\AA\, for a 1\frac.0 slit. At each position
we first took a short 5 min exposure followed by one or two longer 15
min exposures. The instrument response was derived thanks to
observation of the calibration stars LTT\,1020, LTT\,1788, EG\,21.  \\

The seeing varied from 0\frac.45 to 0\frac.53, allowing us to obtain
un-contaminated spectra.  The identification of the stars along the
slits was based on sketches drawn during the observations. Since the
targets lie in crowded regions, some ambiguous identifications
required the development of a small \texttt{IRAF} task, using the
position angle information in the FITS headers. First, each spectrum
was integrated along the dispersion axis, the result being stored into
a two-pixel wide strip, which is close to the actual size of the slit.
Then, the position angle and the pixel-arcsec correspondence were used
to calculate the rotation matrix for the World Coordinate System.
This allowed creation of a slit chart, an $\alpha$-$\delta$ calibrated
two-dimensional image containing accurate slit orientations.
Displaying simultaneously the slit chart besides the SUSI images and
using the WCS correlations it was possible to accurately check the
identity of the star in the slit. \\

\section{Photometry results}
 
\subsection{Components of the clusters}

Fig.~\ref{global} presents  the two tight clusters R\,127 and R\,128 and their
immediately surrounding field (Sect.\,\ref{sec:fieldstars}). 
The central image restored  by deconvolution is  displayed
in Fig.~\ref{dec}, while Tables \ref{photor127}, \ref{photor128} 
and \ref{photobrightest} list the results of the photometry.\\

We show that R\,127 is composed of at least 14 stars. The brightest
component, \#7 with $y=9.37$ and $(b-y)=0.23$ mag, must be the real
LBV (Sect. \ref{sec:r127cluster}). 
The closest components to \#7, stars \#8 and \#3
having $y=14.39$ and 12.12 mag respectively, 
lie at 1\frac.5 and 3\frac.3 from
the main component.  Star \#3 turns out to be very interesting as shown
by its spectrum (Fig.~7).  R\,127 was the target of coronographic
imaging through a broad-band $R$ filter and also a narrow-band
\ha\,+\nb\, filter by \citet{clampin} and \citet{notaIAU143}. 
They detect star \#3 which they call R\,127B.  Those observations also
show the presence of a faint bipolar nebulosity that fails to appear
on our frames. On the other hand, their use of a 2\frac.7-diameter
masking disk centered on R\,127 forbids any detection of the companion
identified as star
\#8. Even star \#10 remains undetected, possibly
due to low S/N ratio. The existence of stars \#8 and \#10 is however
confirmed by an unpublished $R$ frame which we obtained in 1991 with
NTT+SUSI under better seeing conditions.  More recently, R\,127 has
been observed using WFPC2 on {\it HST} (archive data, GO\,6540,
P.I. Schulte-Ladbeck). Again, stars \#8 and \#10 are visible, as well
as several fainter spots, but the combination of the diffraction
pattern, saturation by star \#7, and the presence of nebulosity makes
these spots difficult to identify. \\

R\,128 is resolved into at least 33 stars. It is therefore more
populated than its neighboring cluster R\,127.  The main component,
\#37 with $y= 10.82$ mag, should be considered as the star 
usually referred to as R\,128, since the  next brightest star of
the cluster, \#24 with $y=13.16$ mag, is much weaker. \\

Fig.~\ref{cm}\, shows the color-magnitude (C-M) diagrams of the R\,127
and R\,128 clusters.  The position of the brightest star of the
sample, \#7 ($y=9.37$, $(b-y)= 0.23$ mag), is confirmed by independent
photometric results.  In fact the light curve of R\,127 from 1985 to
2002 (Stahl, private communication) shows that the LBV attained a
brightness of $V=9.35 \pm 0.2$ at the end of the year 1997. Similarly,
the correlation between the magnitude and color of R\,127
\citep{spoon} indicates a color of $(b-y)=0.23\pm 0.05$ mag
corresponding to a visual brightness of $y=9.36$ mag.  \\

The C-M diagram shows a main sequence centered on $(b-y)$\,\ab\,0.00
mag for both clusters.  The intrinsic colors of the O type stars vary
from --0.12 to --0.15 mag for $T_{eff}$ ranging from 30\,000 to
50\,000 degrees and {\it log g} = 3.50 to 5.00 \citep{Relyea78}.
Taking  (b -- y)$_{0}$ = --0.14 mag as the average value and
using the conversion relation $E(b-y) = 0.67 E(B-V) + 0.02$
\citep{kaltcheva}, we find $E(B-V)= 0.18$ mag ($A_{V} = 0.6$ mag) for
the average reddening of clusters R\,127/R\,128.  
The reddest components of the R\,127 cluster are \#8
and \#9, which we will discuss in Sect. 5. The R\,128 cluster
also shows a number of red stars.  \\

The absolute magnitudes, $M_{V}$, listed in Tables 1, 2, and 3 were
derived assuming equality between the Str\"omgren $y$ and Johnson $V$
magnitudes. Taking a distance modulus of 18.6 mag for the LMC
\citep{whitelock}, we used the above-mentioned relation to correct
for the reddening of each star. The corresponding luminosities used in
Sect. 5 were obtained from the bolometric corrections given by
\citet{Vacca96} for O type stars. There is now general agreement that
\citet{Vacca96} overestimate the temperatures and luminosities
\citep{martins}.  However, their calibration is adequate for our purpose 
in this paper, especially since we use their bolometric corrections.  
For B types and the LBV R\,127, which shows a
(peculiar) A type supergiant spectrum, we used the calibration by
\citet{fitz90}. \\

\subsection{Field stars}

\label{sec:fieldstars}

Photometry was obtained for 244 field stars lying in the direction of
the R\,127 and R\,128 clusters. Among them those brighter than $y=16$ mag
are listed in Table\,\ref{photobrightest}.  Stars \#283 and
\#284, situated out of our images, appear in this table because they 
happened to fall on the spectrograph slit and therefore have  
spectral classifications. The brightest stars of this sample are \#61, \#70 
(more commonly known as Sk\,$-69^{\circ}$\,217  and Sk\,$-69^{\circ}$\,218)
and then \#290 and \#113. The C-M diagram for the whole sample of 
the field stars is displayed in Fig.~\ref{cm}, and their detailed 
photometry is available upon request.\\

\begin{table*}[htb]
\caption[]{Photometry of the R\,127 components
\label{photor127}}
\begin{tabular}[h]{c c c c c c c p{.3\linewidth}}
\hline
Star &  $\alpha$ & $\delta$ & Str\"omgren $y$  &  $(b-y)$ & $M_V$& Spectral type & Notes\\
     &  (2000.0) &  (2000.0 &  (mag)           &    (mag) & (mag)& & \\
\hline	 
  1 & 5:36:42.8 & --69:29:58.3	& 17.75 & --0.08 & 	&  & \\	
  2 & 5:36:43.0 & --69:29:56.4	& 15.91 & --0.08 & 	&  & \\	
  3 & 5:36:43.2 & --69:29:45.6	& 12.12 &   0.02 & --7.27	& B0 Ia (N wk) & Fig.~7\\	
  4 & 5:36:43.4 & --69:29:56.0	& 16.12 &   0.01 & 	&  & \\	
  5 & 5:36:43.5 & --69:29:54.7	& 15.03 & --0.04 & 	&  & \\	
  6 & 5:36:43.6 & --69:29:51.5	& 16.78 &   0.10 & 	&  & \\	
  7 & 5:36:43.7 & --69:29:47.4	&  9.37 &   0.23 & --10.39 &Peculiar A supergiant  &  Fig.~\ref{r127sp}. The  LBV object R\,127  \\
  8 & 5:36:43.8 & --69:29:46.2	& 14.39 &   0.35 & 	&  & \\	
  9 & 5:36:43.9 & --69:29:51.3	& 15.78 &   0.24 & 	&  & \\	
 10 & 5:36:44.0 & --69:29:44.5	& 17.15 &   0.13 & 	&  & \\	
 11 & 5:36:44.6 & --69:29:49.6	& 15.81 &   0.02 & 	&  & \\	
 12 & 5:36:44.6 & --69:29:52.2	& 17.37 & --0.11 & 	&  & \\	
 13 & 5:36:44.6 & --69:29:46.0	& --    &    --  & 	&  & \\	
 14 & 5:36:44.8 & --69:29:57.3	& 18.87 &    --  & 	&  & \\
\hline
\end{tabular}
\end{table*}

\begin{table*}[htb]
\caption[]{Photometry of the R\,128 components
\label{photor128}}
\begin{tabular}[h]{c c c c c c c p{.3\linewidth}}
\hline
Star &  $\alpha$ & $\delta$ &  Str\"omgren $y$ & $(b-y)$ & $M_V$ & Spectral type & Notes\\
     &  (2000.0) & (2000.0) &     (mag)        &   (mag) & (mag) & &\\
\hline	 
 15 & 5:36:45.7 & --69:29:55.5 & 16.92 & 0.29 &    	&         &  \\		
 16 & 5:36:45.9 & --69:29:58.9 & 14.94 & 0.02 &	--4.45  & O9.7 II & Fig.~6\\	
 17 & 5:36:45.9 & --69:30:00.7 & 18.09 & 0.00 &	   	&         &\\	
 18 & 5:36:46.3 & --69:30:00.2 & 18.07 &  --  &	   	&         &\\	
 19 & 5:36:46.3 & --69:29:52.8 & 18.17 & 0.16 &	   	&         &\\	
 20 & 5:36:46.3 & --69:29:56.7 & 16.30 & --0.02 & 	&         &\\
 21 & 5:36:46.4 & --69:30:02.6 & 17.22 & --0.21 &  	&	  &\\	
 22 & 5:36:46.5 & --69:30:01.1	& 14.63 & --0.02 & --4.58	  & O9 III & Fig.~5 \\	
 23 & 5:36:46.5 & --69:30:04.4	& 15.50 & --0.03 & --3.66	  & O9 Ib  & Fig.~5 \\	
 24 & 5:36:46.5 & --69:29:56.1 & 13.16 & --0.04 &  --5.95	  & O8.5 II &Fig.~4 \\	
 25 & 5:36:46.5 & --69:29:58.3	& 16.50 &   0.00 & 	&	  &  \\	
 26 & 5:36:46.5 & --69:30:02.8	& 15.29 & --0.03 & --3.87	  & O9.7 II & Fig.~6\\	
 27 & 5:36:46.5 & --69:29:51.8	& 17.12 &   0.28 & 	& 	  & \\	
 28 & 5:36:46.7 & --69:29:59.8	& 14.59 & --0.04 & --4.52	  & O9.5 III & Fig.~6 \\	
 29 & 5:36:46.7 & --69:30:01.3	& 18.14 &   0.22 & 	&         & \\	
 30 & 5:36:46.8 & --69:30:04.4	& 17.33 &   0.06 & 	&         & \\	
 31 & 5:36:46.8 & --69:29:58.9	& 16.99 &   0.10 & 	&         & \\	
 32 & 5:36:46.9 & --69:29:57.7	& 15.67 & --0.06 & --3.35	  & O9.5 III & Fig.~6\\	
 33 & 5:36:47.0 & --69:30:00.2	& 15.82 &   0.05 & 	&	  & \\	
 34 & 5:36:47.0 & --69:29:49.1	& 17.37 &   0.23 & 	&	  & \\	
 35 & 5:36:47.1 & --69:29:59.3	& 15.46 & --0.04 & --3.65	  & B0 III & Fig.~7. Possible 
contamination with \#33 \\			   
 36 & 5:36:47.1 & --69:29:56.3	& 17.47 &   0.30 & 	&	  & \\	
 37 & 5:36:47.2 & --69:29:52.2	& 10.82 &   0.05 & --8.71	  & B1.5 Ia (N wk)&Fig.~7.  The ``true'' R\,128\\
 38 & 5:36:47.2 & --69:30:03.9	& 19.09 &   --   & 	&	  & \\	
 39 & 5:36:47.4 & --69:29:58.0	& 15.67 & --0.01 & --3.58	  & B0 III & Fig.~7\\	
 40 & 5:36:47.4 & --69:30:01.0	& 19.11 &   --   & 	&	  & \\	
 41 & 5:36:47.4 & --69:29:59.7	& 19.09 &   --   & 	&	  & \\	
 42 & 5:36:47.7 & --69:29:58.4	& 17.72 & --0.03 & 	&	  & \\	
 43 & 5:36:47.7 & --69:29:51.8	& 15.83 & --0.07 & 	&	  & \\	
 44 & 5:36:47.9 & --69:29:59.7	& 18.41 & --0.05 & 	&	  & \\	
 45 & 5:36:48.0 & --69:29:56.0	& 13.37 & --0.05 & --5.70	  & O9 II & Fig.~5 \\	
 46 & 5:36:48.1 & --69:29:51.3	& 16.37 &   0.01 & 	&         & \\
 47 & 5:36:48.2 & --69:29:53.6	& 16.36 & --0.07 & 	&	  & \\	
\hline
\end{tabular}
\end{table*}

\begin{table*}[htb]
\caption[]{Photometry of the brightest field stars
\label{photobrightest}}
\begin{tabular}[h]{c c c c  c c c p{.3\linewidth}}
\hline
Star & $\alpha$ & $\delta$ & Str\"omgren $y$ &  $(b-y)$ & $M_V$ & Spectral types & Notes\\
     &  (2000.0) & (2000.0)&     (mag)       &    (mag) & (mag) & & \\
\hline	 
61 & 5:36:37.1 & --69:29:18.4  & 11.94 & --0.08 &	--6.99	& 
O7 Iaf & Fig~4. \object{Sk--69$^{\circ}$\,217}. Apparent weak emission line
redward of  H-delta is an artifact
\\ 
 65 & 5:36:37.8 & --69:30:16.7 & 15.90 & --0.05 &		& & \\
 67 & 5:36:37.7 & --69:29:13.0 & 15.51 & --0.03 &		& & \\
 69 & 5:36:37.8 & --69:29:40.8 & 15.03 & --0.04 &		& & \\
70 & 5:36:37.8 & --69:29:33.2 & 11.92 & --0.03 & 	--7.24	&
O8 Iab(f) &  Fig.~4. \object{Sk--69$^{\circ}$\,218}. Apparent weak emission   
line near \heb\,\,$\lambda4686$ is blueward of that wavelength
\\
 97 & 5:36:39.9 & --69:30:07.3 & 15.86 &   0.90 &		& & \\
100 & 5:36:40.0 & --69:29:31.6 & 14.76 &  0.01 &		& & \\
109 & 5:36:40.6 & --69:30:09.9 & 15.53 &  0.53 &		& & \\
112 & 5:36:40.8 & --69:29:29.8 & 15.82 & --0.05 &		& & \\
113 & 5:36:40.8 & --69:29:53.1 & 14.00 &  0.00 &	--5.30	& O9.7 II & Fig.~6 \\
159 & 5:36:43.9 & --69:29:09.8 & 15.81 & --0.08 &		& & \\
160 & 5:36:43.9 & --69:30:02.3 & 15.88 & --0.08 &		& & \\
161 & 5:36:44.0 & --69:29:26.8 & 15.84 &   0.37 &		& & \\
162 & 5:36:44.1 & --69:30:05.1 & 15.45 & --0.10 &		& & \\
211 & 5:36:46.8 & --69:29:13.9 & 15.41 & --0.09 &		& & \\
217 & 5:36:47.2 & --69:30:29.0 & 15.26 & --0.03 &		& & \\
219 & 5:36:47.4 & --69:30:27.1 & 14.63 & --0.08 &		& & \\
222 & 5:36:47.5 & --69:30:09.4 & 15.58 & --0.10 &		& & \\
249 & 5:36:49.3 & --69:29:30.8 & 15.37 & --0.11 &		& & \\
265 & 5:36:50.6 & --69:29:57.4 & 15.64 &   0.07 &		& & \\
268 & 5:36:50.8 & --69:30:16.7 & 15.52 &   0.08 &		& & \\
281 & 5:36:56.8 & --69:29:27.3 & 15.15 & --0.31 & 		&
O9.7: &  Fig.~6. \heb\,\,$\lambda$\,4686 absorption is weak indicating high luminosity, 
but \sid\,\,  is also weak indicating the opposite. This discrepancy 
is likely due to noise rather than a real peculiarity.\\
283 & 5:36:34.3 & --69:30:41.0 &  --    &  --     & 	   & B0 III & Fig.~7 \\
284 & 5:36:31.0 & --69:30:53   &  --    &  --     & 	   & B1.5   & Fig.~7 \\
285 & 5:36:51.8 & --69:30:24.3 & 14.03 & -- &		& & \\
286 & 5:36:52.6 & --69:30:22.8 & 15.30 & -- &		& & \\
287 & 5:36:53.2 & --69:29:04.9 & 14.72 & -- &		& & \\
288 & 5:36:53.2 & --69:28:40.8 & 14.51 & -- &		& &\\
289 & 5:36:53.8 & --69:30:23.0 & 14.83 & -- &	 	& &\\
290 & 5:36:54.2 & --69:28:58.2 & 13.61 & -- &		& &\\
291 & 5:36:54.3 & --69:28:43.6 & 15.76 & -- &	 	& &\\
\hline
\end{tabular}
\end{table*}

\section{Spectral types}
\label{classif}

The spectral classification of the OB stars was performed with respect
to the digital atlas of \citet{wal-fitz}.  No
classification standards with the current observational setup were
available, so there may be a slightly greater uncertainty in the
(lower) luminosity classes than is usual in this work, but it does not
exceed one class and may well be smaller.\\

\parindent=0cm

\subsection{R\,127 cluster}
\label{sec:r127cluster}

Spectrograms of the two brightest stars, \#3 and \#7, are available.
These are the ``preceding'' (western) and ``following'' (eastern)
components of R127 (HDE\,269858), respectively, as denoted by \citet{feast}, 
and we confirm that the eastern component, \#7, is the
peculiar star.  The two components had similar apparent magnitudes
when observed by \cite{feast}, but as discussed in the Introduction,
\#7 is currently in an extended LBV  phase and is much brighter. Our
observation of its spectrum is illustrated in Fig.~3.  It is
representative of the peculiar A supergiant LBV maximum phase,
dominated by emission lines of hydrogen and singly ionized heavy
metals (e.g., \feb, [\feb], \tib, \crb); see \citet{wolf88} for a
detailed presentation of this kind of spectrum at high resolution. \\

\parindent=0.5cm

We believe that our spectrogram of star \#3 (Fig.~7) is the first
spatially resolved observation of its spectrum.  It has an
interesting, very well defined B0~Ia type with deficient nitrogen
(cf., e.g., the extreme weakness of the $\lambda4097$ absorption in
the blue wing of H$\delta$, which is near maximum strength in a normal
spectrum of this type).  The latter characteristic indicates that this
star either is a relatively un-evolved supergiant, or was a slow
rotator on the main sequence (e.g., \citet{walborn2000} and references
therein). \\

\parindent=0.0cm
\subsection{R\,128 cluster}

We have spectroscopy for 11  stars in this compact cluster, which
is shown in Figs.~4--7.  Most of them are late-O or early-B giants
(i.e., the main sequence was not reached spectroscopically), but one
(\#23, Fig.~5) is classified O9~Ib, and \#37 is R\,128 itself, the
brightest star in the cluster, which we classify B1.5~Ia (N weak).
R\,128 was classified B1~Ia by \cite{fitz88} and B2~Ia (N weak) by
\citet{fitz91}.  We prefer the intermediate type on the basis of
the strength of \sid\, $\lambda4089$ and concur on the anomalous
deficiency of nitrogen; e.g., \nb\, $\lambda3995$ reaches its maximum
strength at B2~Ia and would be comparable to \hea\,\,$\lambda4026$ in a
normal spectrum.  Current explanations of such anomalies are as cited
above for star \#3 in the R\,127 cluster. \\

\subsection{Field stars}

Finally, 6  stars in the surrounding field were also observed
spectroscopically.  The most interesting are the relatively bright
stars \#61 (Sk\,$-69^{\circ}\,217$) and \#70 (Sk\,$-69^{\circ}\,218$) shown
in Fig.~4.  The Of nature of Sk\,$-69^{\circ}217$ was discovered by
\citet{walbornIAU143}  and the spectral classification of O7~Iaf from the
present, higher quality material is in perfect agreement with the
earlier result; this is the hottest star in the present sample.  On
the other hand, the earlier classification of ON9~Ib for
Sk\,$-69^{\circ}\,218$ is here revised to O8~Iab(f), with no
nitrogen/carbon anomaly.  This discrepancy can be understood in terms
of the lower quality of the earlier spectrogram, which led to a later
type at which the N/C line strengths would be anomalous, but they are
not at the earlier type derived here.  The recently identified \sd\,
$\lambda\lambda\,4486$, 4504 emission lines \citep{werner} are
prominent in both of these spectra.\\

\section{Discussion}

The brightest members of the R\,127/R\,128 clusters are massive
evolved stars as indicated by their spectral types. There are,
however, stars residing still on the main sequence (Fig.\,\ref{cm}),
but they have been missed in our spectroscopy due to their relatively
low brightness. In order to look into the evolutionary states of the
clusters' stars, we have used the theoretical models of
\citet{meynet94} for a metallicity of z\,=\,0.004 and high mass loss
rates. Fig.\,\ref{hrd} presents the relevant isochrones and
evolutionary tracks on which are overlaid the stellar positions. The
filled dots are based on the physical parameters given by
\citet{Vacca96}'s calibration for the derived spectral types. 
The crosses indicate the corresponding luminosities obtained from our
photometry. We see that for a number of stars the observed
luminosities are much smaller than those predicted by
\citet{Vacca96}'s calibration. We will briefly discuss  this
problem below. \\

\parindent=0.5cm

The positions of the filled dots suggest that the hottest stars
of the sample, \#61 and \#70, are \ab\,3 Myr old, and have evolved
from initial masses of  \ab\,80\,\sm\, into late-type O supergiants.
Likewise, the oldest stars, \#39 and \#35, are B0 supergiants with an
age of \ab\,6 Myr and a ZAMS mass of \ab\,25\,\sm. 
 It seems therefore that in the R\,127/R\,128 region lower
mass stars have formed prior to massive ones. Moreover, the positions
of the crosses, taken at face value, suggest masses as low as 
\ab\,15\,\sm\, with ages up to \ab\,8 Myr.   \\

The LBV member of  the sample, star \#7, does not appear in the H-R
diagram since its relatively low effective temperature puts it outside
the plot. It is well-known that LBV stars change their
spectral type, temperature  and radius, becoming cooler and larger
during the visual maximum phase. However, their luminosity remains
approximately constant, with {\it log} $L/L_{\odot}$\,=\,6.1 for
R\,127, as reported by several workers \citep{stahl83, wolf89, lam98,
gend01}. An effective temperature of {\it log} $T_{eff}$\,=\,3.954 is
calculated by \citet{lam98} for R\,127 corresponding to its phase of
visual maximum. \\

Star \#7 is the most evolved member of the R\,127/R\,128 clusters,
presumably because it was the most massive star of the group. An
initial mass of \ab\,85\,\sm\, can be attributed to the LBV progenitor
from Fig.~\ref{hrd}. A present mass of 46$^{+17}_{-3}$ \sm\, is
derived for the LBV by \citet{lam98}. This means that star \#7 has
lost \ab\,45\% of its mass since its formation \ab\,3 Myr ago, 
corresponding to a constant mass loss rate of \mdot\,=\,1.3\,\x\,10$^{-5}$
\sm\,yr$^{-1}$. However, mass loss is not a regular process
and LBVs undergo drastic episodes during which they lose mass
dramatically. For example, model calculations by \citet{langer94} show
that during the first 1.5 Myr of life of a 60\,\sm\, star the mass loss
rate is constant, \ab\,5\,\x\,10$^{-6}$ \sm\,yr$^{-1}$. Then it
becomes much stronger, \ab\,3\,\x\,10$^{-5}$ \sm\,yr$^{-1}$, during
the next 2 Myr.  The peak of mass loss during the LBV phase reaches
\ab\,3\,\x\,10$^{-3}$
\sm\,yr$^{-1}$ when the star has an age of \ab\,3.4 Myr.
On the other hand, association in a close binary system has been
suggested for accelerating the evolution of massive stars and creating
the LBV phenomenon (\cite{tutukov, gallagher, humph94}).  However, so
far no observational support exists for this LBV being a binary
system.  \\

Fig.\,\ref{hrd} shows a discrepancy between the luminosities suggested
by the spectral type calibration schemes \citep{Vacca96} and those
derived from the photometry.  While the smaller deviations, like those
of stars \#61, \#70, \#3, \#45, etc., can be attributed to measurement
uncertainties, a discrepancy as large as that for star \#23 called for
verification. This trend was first interpreted as due to a sort of
bias. However, despite our efforts, no systematic errors were found in
our photometry. In fact, our examination of several published papers on
other regions showed that $\Delta M_V$ (the difference between the
absolute magnitudes derived from spectral types and those from
photometry) increases with magnitude, although with different extents 
\citep{fitz88, walborn97, walborn99, parker92, massey93}.
 Such trends are expected from the defining relations, if there
are dispersions in the actual absolute magnitudes or distances, or
systematic effects in the spectral classification, absolute-magnitude
calibration, or photometry.  Uncorrected differential extinction or
relative errors in the calibration will broaden the distribution about
the trends. \\

Aperture photometry observations previously performed on R\,127 used
diaphragms of 10\,\frac\, to 15\frac\, in diameter. As a result, the
global magnitude of a small group of stars was mistaken for  that of
a single source. Assuming that we can simulate such a measurement by
co-adding the stars found inside a circle of 10\,\frac\, centered on
star \#7 (which are stars \#8, \#3, \#10, \#9 and \#6), we find that
the measured $y$ magnitude would be 9.27 instead of 9.37. This small
difference shows that the contribution of the weak companions
identified in this work is not significant for the luminosity of the
main object. \\

As for the brightest member of the R\,128 cluster, star \#37, it is a
supergiant B1.5 Ia (N wk) with {\it log} $L/L_{\odot}$\,=\,5.76 and
{\it log} $T_{eff}$\,=\,4.24 \citep{gend01} and therefore lies outside
our H-R diagram (Fig.~\ref{hrd}). This star is known to be variable
and has even previously been dubbed as an LBV candidate
 \citep{gend01}.  Indeed an initial mass of \ab\,50\,\sm, 
as derived from Fig.~\ref{hrd}, does not rule out the possibility 
for star \#37 to become an LBV  \citep{maeder_kona, langer94}. \\

As mentioned in Sect. 3.1, some of the cluster components show rather
red colors. In particular, star \#8, the closest neighbor of the LBV,
lying 1\frac.5 away, has the reddest color.  This reminds one of the
LMC transition star R\,84 which has a close evolved component of type
M2 Ia lying \ab\,1\frac.7 from it \citep{mhm97}. Note, however, that
R\,84 is a lower mass object belonging to a B-type cluster.  Anyhow,
presently we cannot determine whether star \#8 is evolved, and the red
color of this star may have a totally different origin.  We are aware
of the color accuracies involved (Sect. 3.1), nevertheless if star \#8
is physically associated with the cluster, and not a sight-line
coincidence, its color could be due to contamination by the ejecta
from the LBV. \citet{clampin} measure a size of 8\frac.0\,\x\,9\frac.0
(1.9\,\x\,2.2 pc) for the circumstellar nebula lying at a position
angle of 165\deg. Even a smaller nebula, as found by
\cite{stahl87}, with a size of   3\frac.2\,\x\,4\frac.4 (0.8\,\x\,1.1 pc), 
can affect the color measurements of star \#8. A larger nebula 
would also explain the red color of star \#9. \\

\parindent=0cm

\section{Concluding remarks}

The decomposition of the LMC compact star clusters R\,127 and R\,128
into at least 14 and 33 components respectively and 
medium-dispersion spectroscopy carried out in very good seeing
conditions have allowed us to study the stellar contents of these
interesting objects. This work is essential  since the R\,127 cluster
harbors the most prominent LBV object in the LMC. We resolve the core
of R\,127 into four components and clearly identify the LBV object,
which shows spectral features typical of a visual maximum phase.
Moreover, we present the first spatially resolved spectrum of the
brightest neighbor of the LBV, star \#3 lying at a separation of
3\frac.3. It turns out to be a supergiant B0 Ia (N wk).\\

\parindent=0.5cm

We also show that the star currently known as R\,128 is in fact the
component number \#37 of an adjacent cluster. A good quality spectrum
of this supergiant leads to a revised classification of B1.5 Ia 
(N wk).\\

The two clusters are composed of evolved massive stars. The oldest
members are  \ab\,6--8 Myr old and the most massive one has an initial
ZAMS mass of  \ab\,80\,\sm. These age estimates are in agreement with
the fact that no \h2\, regions are associated with the clusters.  The
most massive stars have had enough time to disrupt the molecular cloud
and the nebula, and the earliest spectral types are no longer
present. \\

This spectroscopic study misses the main sequence stars of the
clusters, which probably have lower initial masses with respect to the
evolved ones.  It will therefore be interesting to observe them
spectroscopically in order to determine their real status. It will
also be attractive to study the dynamics of the clusters by obtaining
accurate radial velocities of the members.

\begin{acknowledgements}
We are particularly indebted to Dr. Pierre Magain, Institut
d'Astrophysique et de G\'eophysique de l'Universit\'e de Li\`ege,
Belgium, and his group for their warm hospitality offered to
F.M. during his stay in Li\`ege for learning to use the MCS
deconvolution algorithm. Without their patience, availability, and
help the training could not be fruitful. Our thanks go also to
Dr. Fr\'ed\'eric Courbin for discussions and advice at several
opportunities during his trips to Paris.  We are also grateful to
the referee, Dr. A.M. van Genderen, for his
helpful comments.  We would like also to thank Dr. Otmar Stahl,
Landessternwarte, Heidelberg, for discussions and information about
the monitoring of R\,127 and  Dr. Jes\'us Ma\'{\i}z-Apell\'aniz, 
STScI, about the absolute-magnitude trends.
\end{acknowledgements}

\bibliographystyle{aa}
\bibliography{ms3295}

\begin{figure*}
\resizebox{18cm}{!}{\includegraphics{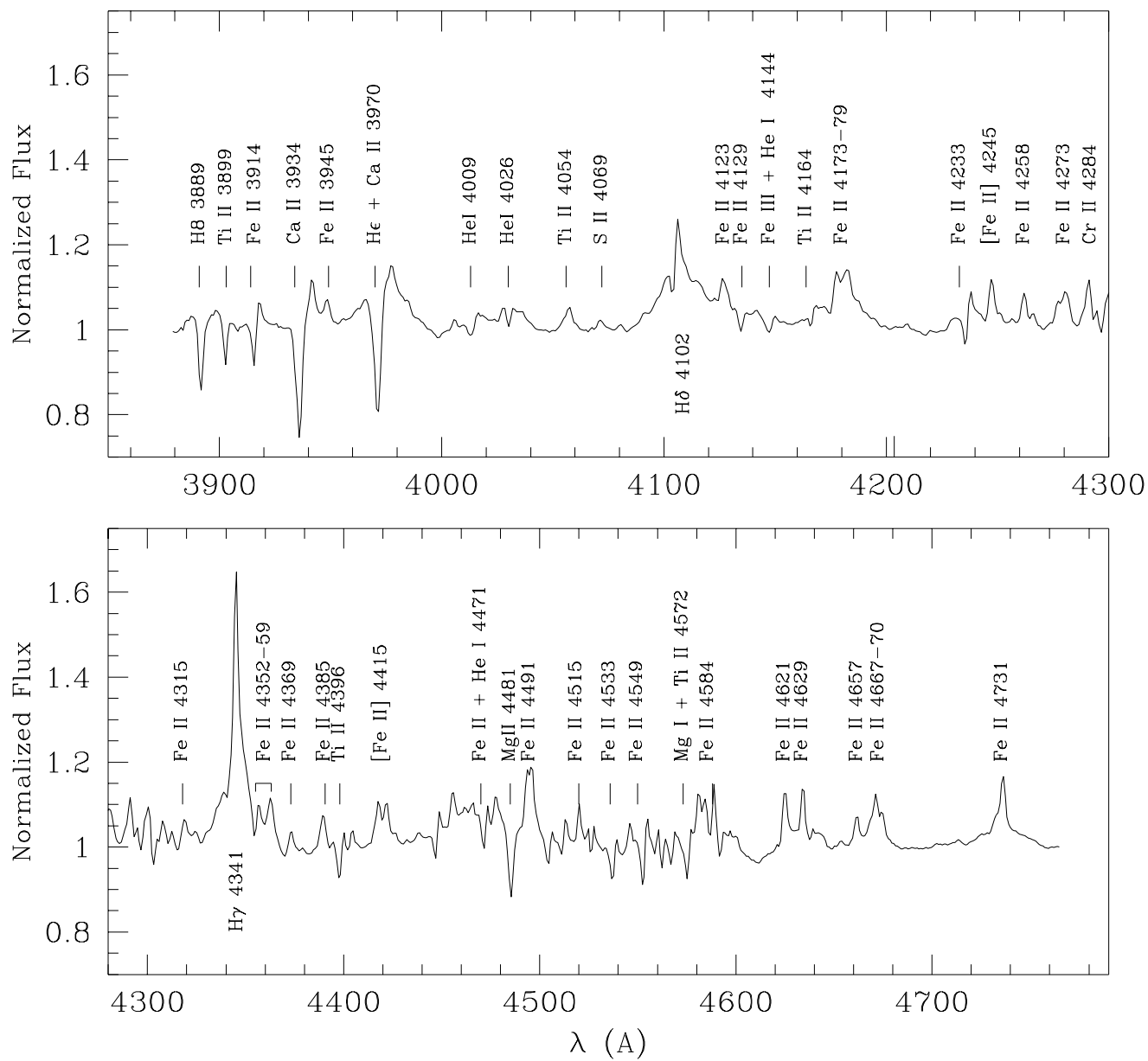}}
\caption{A spectrum of R\,127 obtained in 1997 November using the 
ESO NTT+EMMI with grating \#12.     } 
\label{r127sp}
\end{figure*}

\begin{figure*}
\resizebox{18cm}{23cm}{\includegraphics{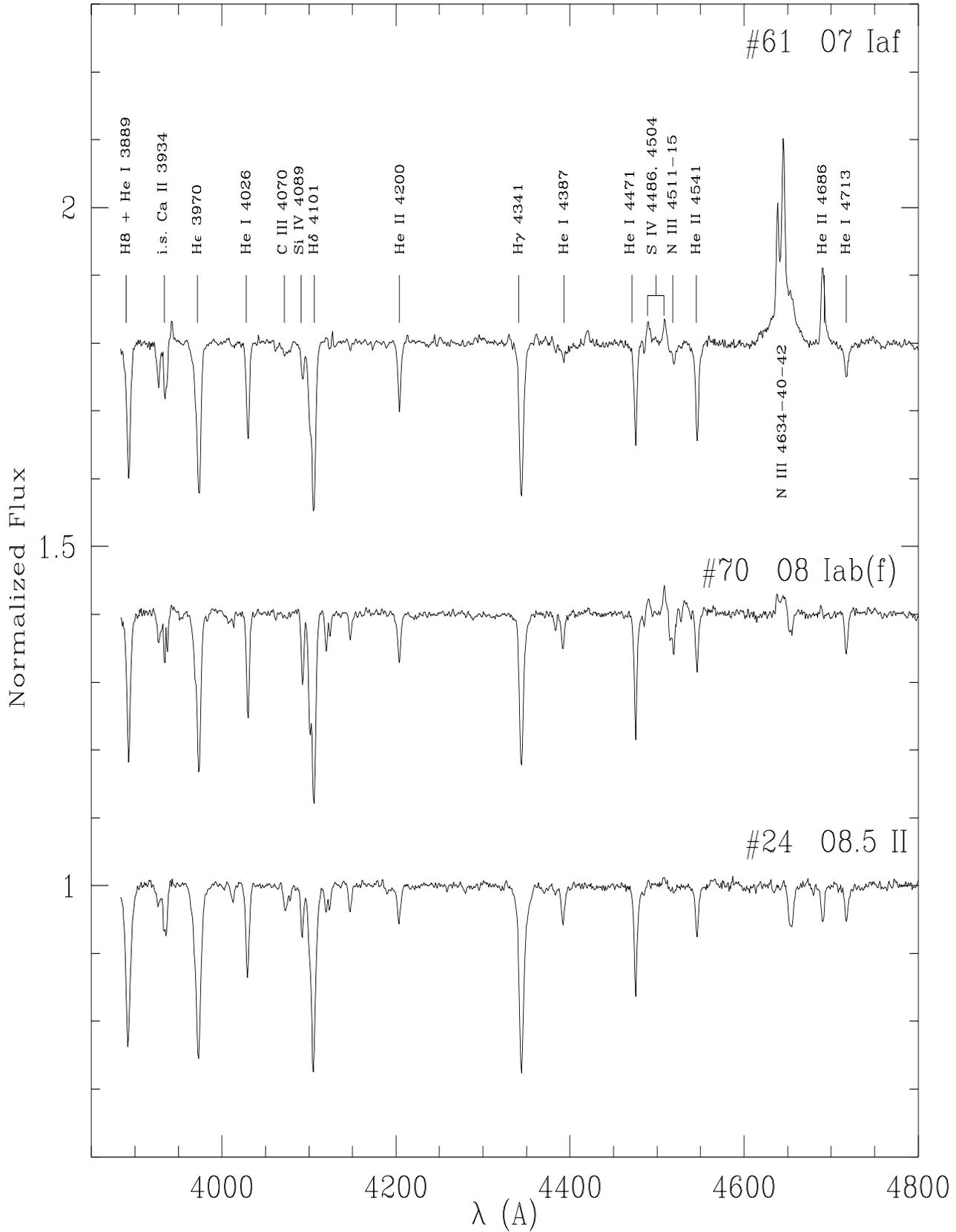}}
\caption{Spectra of the earliest type stars towards R\,127 and R\,128 clusters.} 

\end{figure*}

\begin{figure*}
\resizebox{18cm}{23cm}{\includegraphics{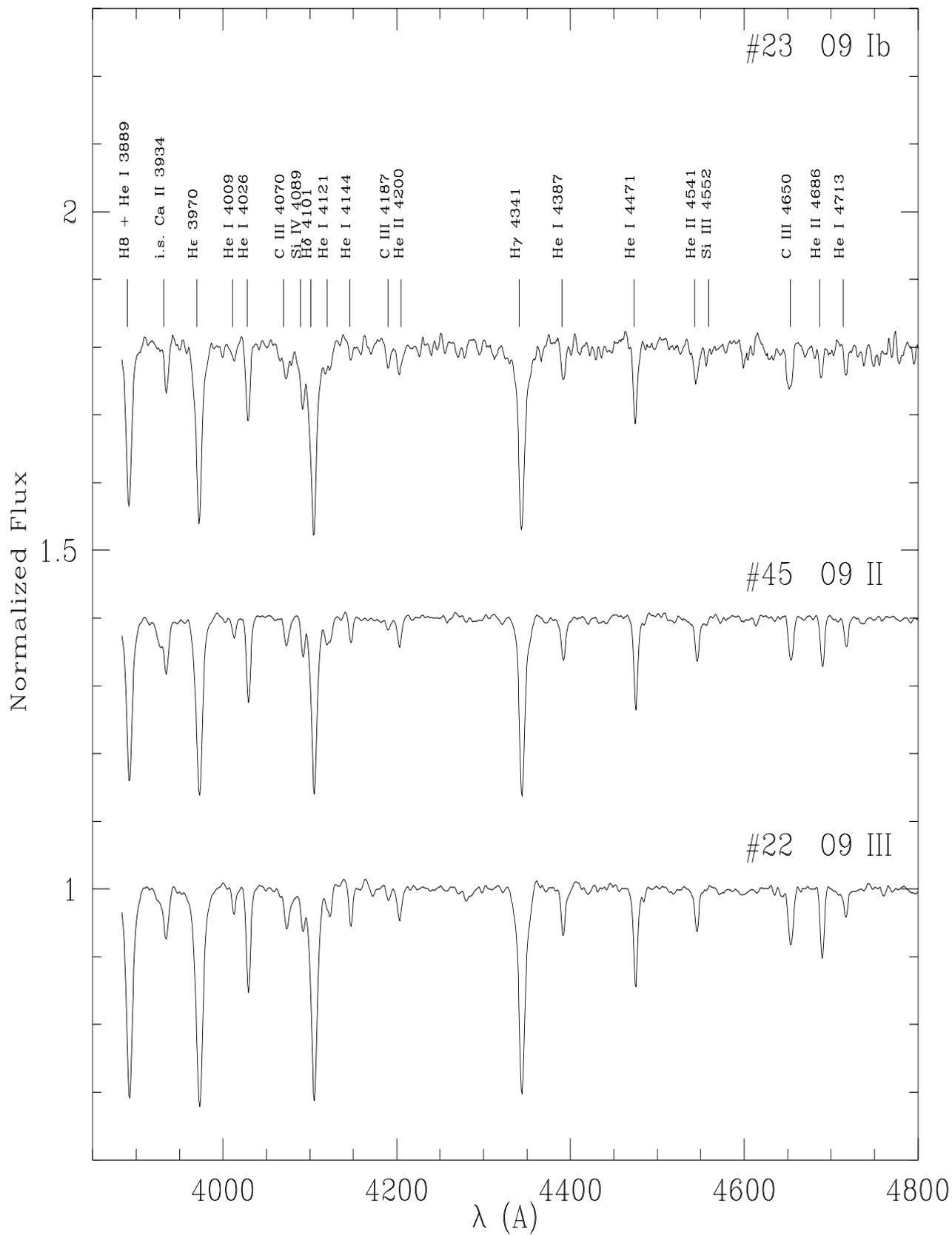}}
\caption{Spectra of O9 type stars observed towards R\,127 and R\,128 clusters. } 
\end{figure*}

\begin{figure*}
\resizebox{18cm}{23cm}{\includegraphics{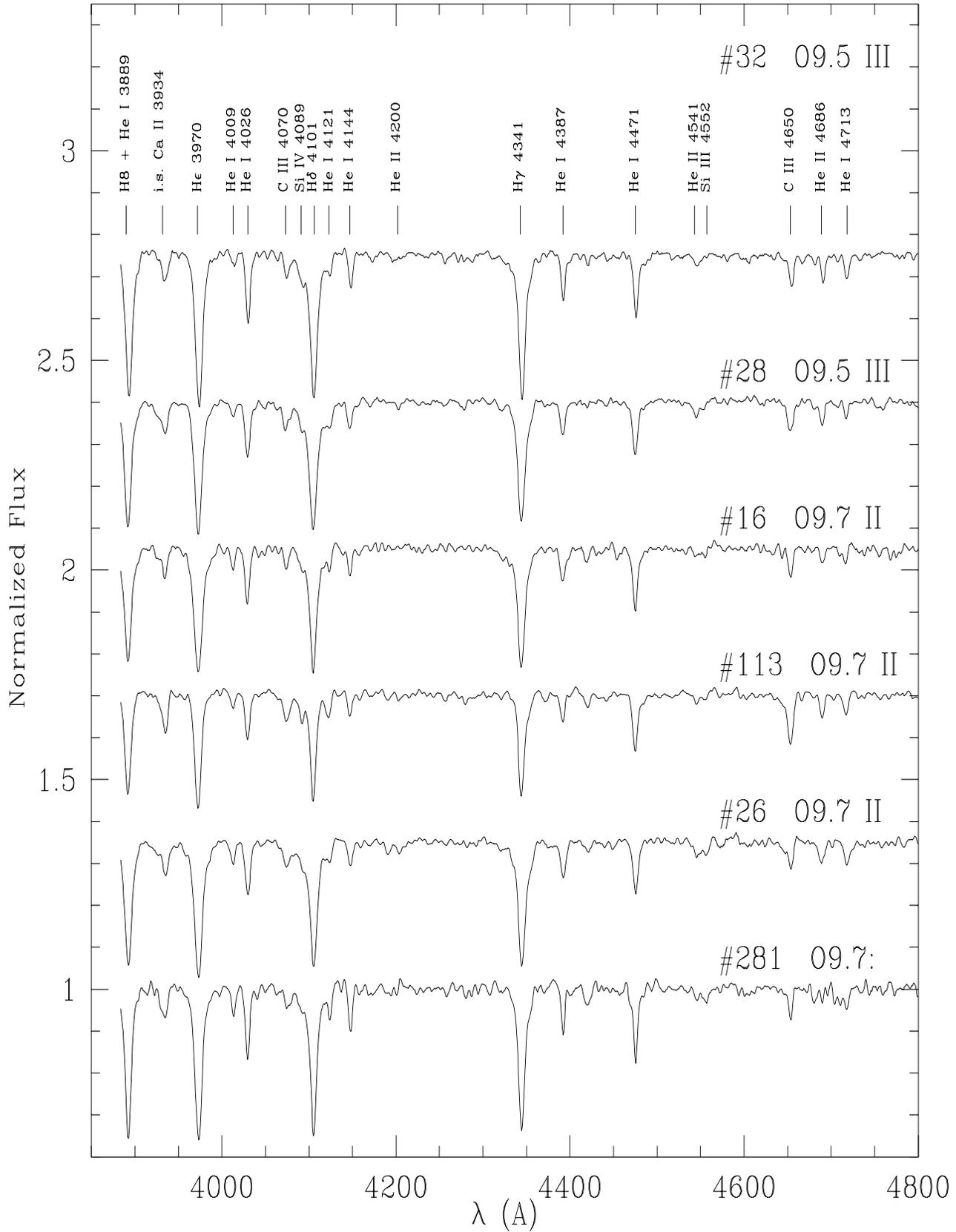}}
\caption{Spectra of the latest O type stars observed towards R\,127 and 
R\,128 clusters. } 
\end{figure*}

\begin{figure*}
\resizebox{18cm}{23cm}{\includegraphics{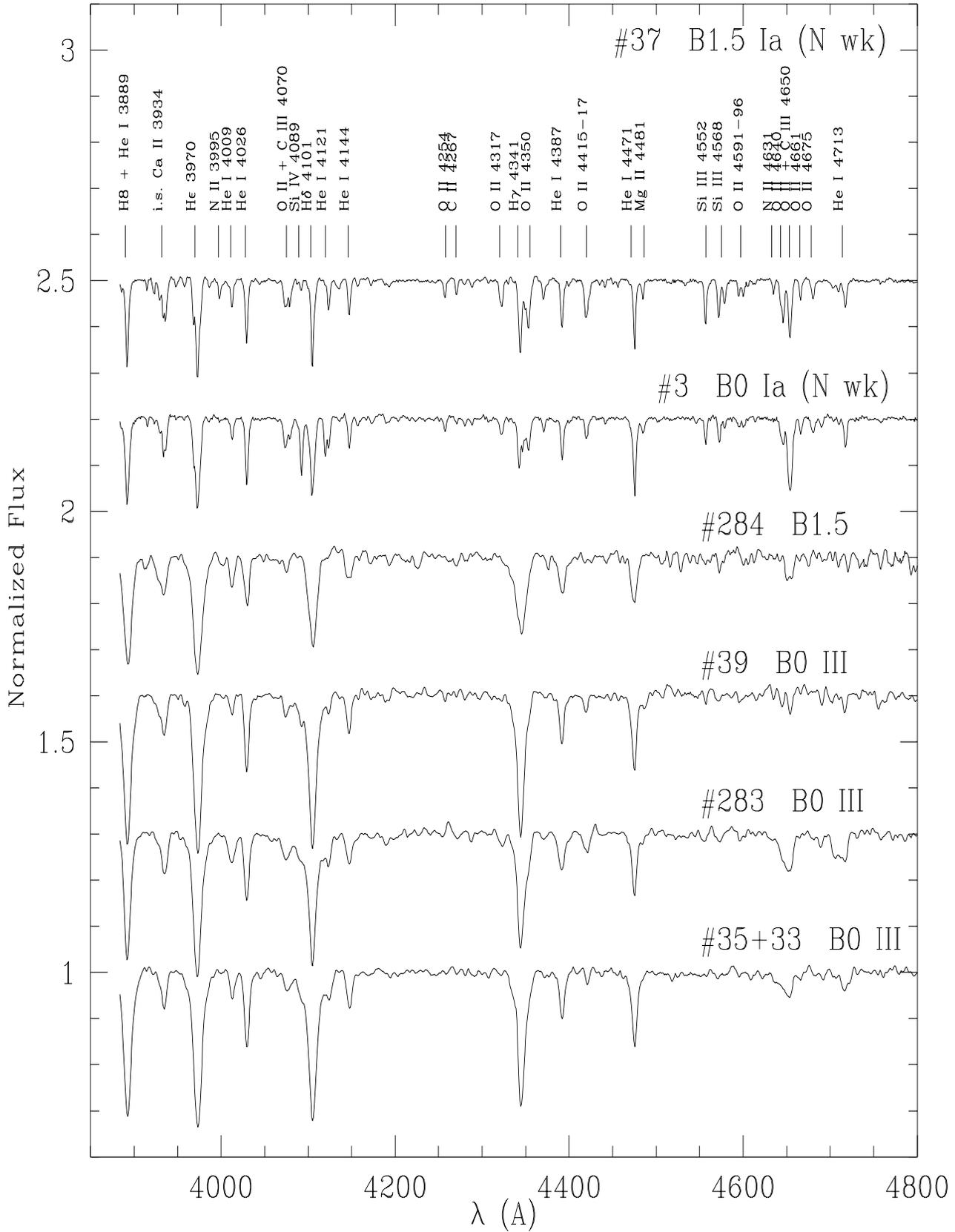}}
\caption{B type stars observed towards R\,127 and R\,128 clusters. } 
%\label{hnt}
\end{figure*}

\begin{figure*}
\resizebox{8.5cm}{!}{\includegraphics{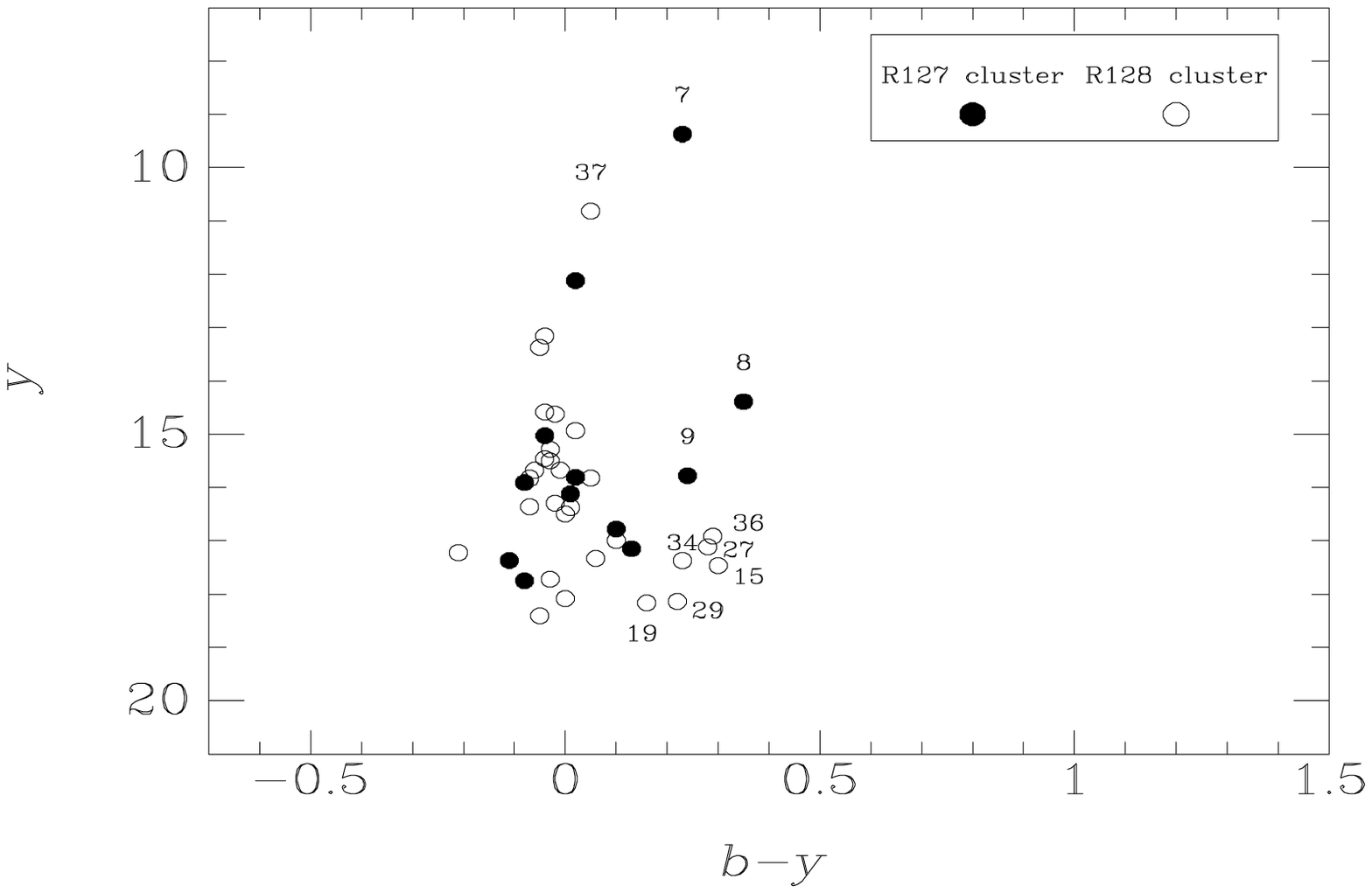}}
\resizebox{8.5cm}{!}{\includegraphics{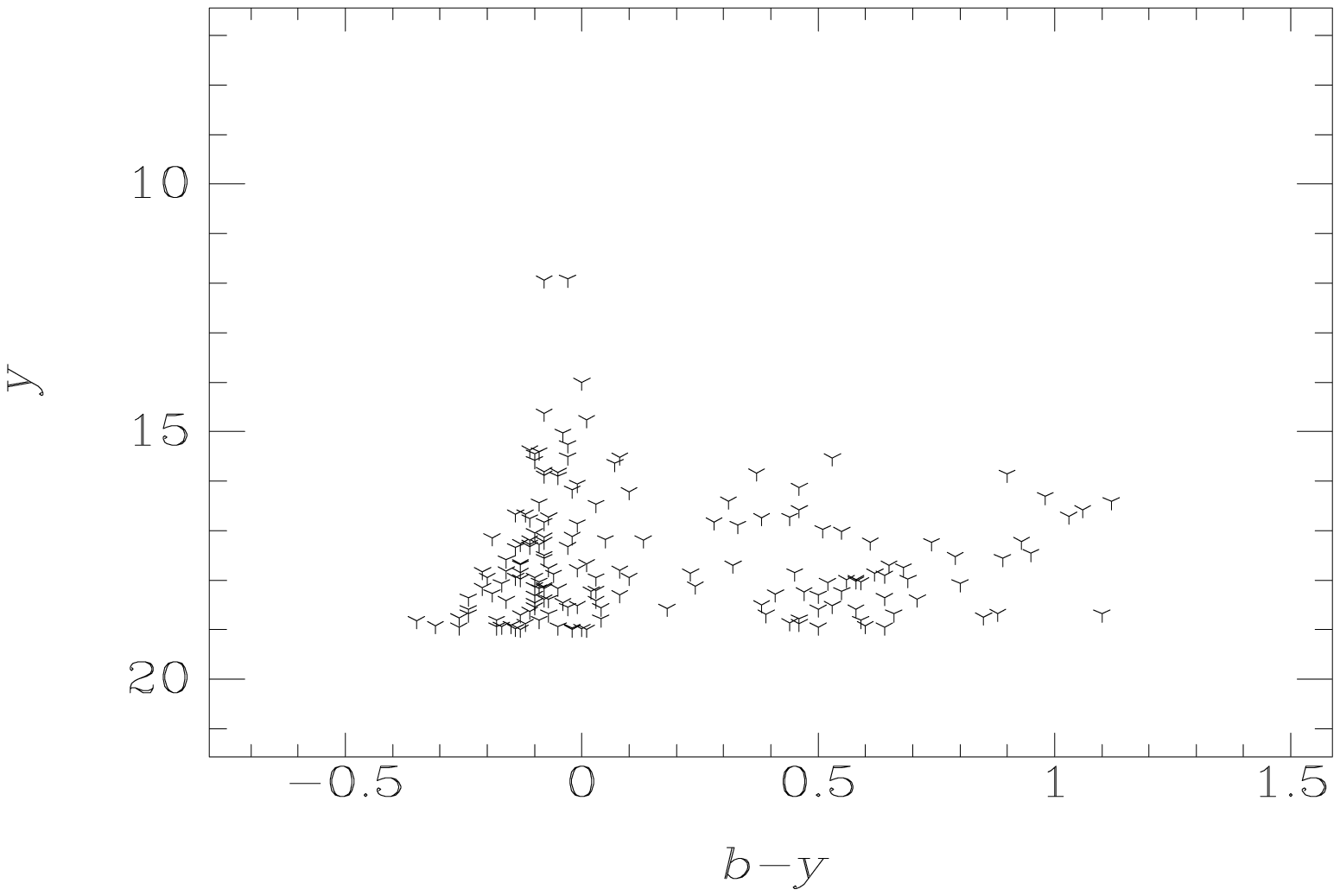}}
\caption{Color-magnitude diagram of stars in R\,127 and R\,128 
  clusters (left panel). Diagram for the field stars around
  R\,127 and R\,128 clusters (right panel). }
\label{cm}
\end{figure*}

\begin{figure*}
\resizebox{18cm}{!}{\includegraphics{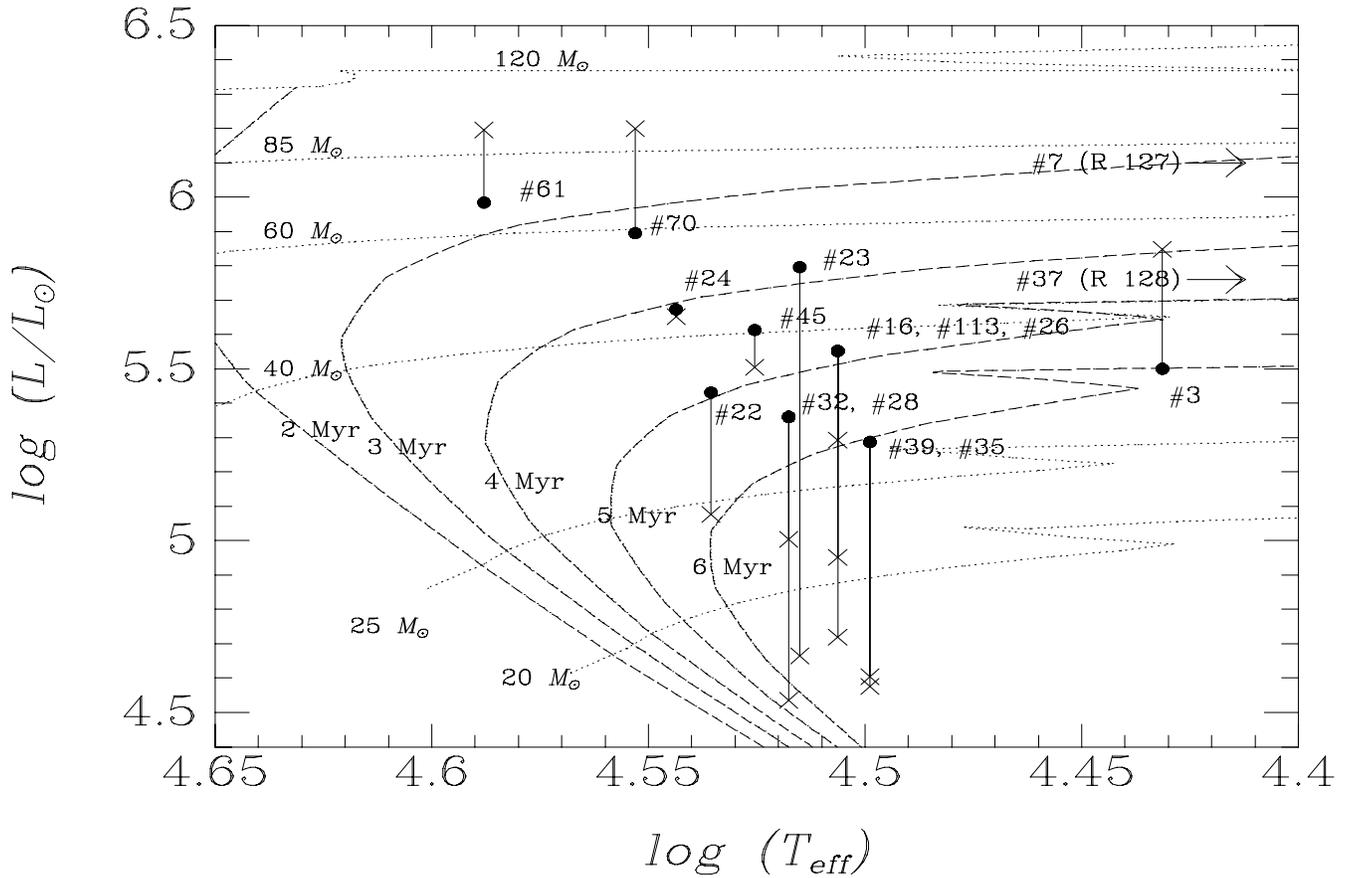}}
\caption{Theoretical H-R diagram (\citep{meynet94} for classified 
stars in the LMC clusters R\,127 and R\,128. The dashed lines indicate
isochrones with ages 2 to 6 Myr and the dotted lines the evolutionary
tracks for masses 20 to 120\,\sm. Filled circles represent the stars
for which spectral classification is available. The positions are
based on the calibration derived by \citet{Vacca96} for O  and early 
B stars and that by \citet{fitz90} for later types. The crosses
represent the luminosities calculated using the photometric results of
this paper.  }
\label{hrd}
\end{figure*}

\end{document}